\title{\LARGE \bf Bayesian sample size determination in basket trials  borrowing information between subsets}

\author{HAIYAN ZHENG$^{1, 2, \ast}$, MICHAEL J. GRAYLING$^2$, PAVEL MOZGUNOV$^1$, \\ THOMAS JAKI$^{1, 3}$, JAMES M. S. WASON$^{2}$ \\[4pt]
\textit{\small $^1$MRC Biostatistics Unit, University of Cambridge, U.K.} \\[-2pt]
\textit{\small $^2$Population Health Sciences Institute, Newcastle University, U.K.} \\[-2pt]
\textit{\small $^3$Department of Mathematics and Statistics, Lancaster University, U.K.} \\[2pt]
$^\ast$Email: {\url{haiyan.zheng@mrc-bsu.cam.ac.uk}}}

\date{ }

\documentclass[11pt]{article}

\usepackage[margin=0.78in]{geometry}
\usepackage{hyperref}
\usepackage{amssymb, amsmath}
\usepackage{graphicx, caption}
\usepackage{booktabs}
\usepackage{multirow}

\usepackage{natbib}

\usepackage{pdfpages}

\begin{document}
\maketitle

\begin{abstract}
Basket trials are increasingly used for the simultaneous evaluation of  a new treatment in various patient subgroups under one overarching protocol. We propose a Bayesian approach to sample size determination in basket trials that permit borrowing of information between commensurate subsets. 
Specifically, we consider a randomised basket trial design where patients are randomly assigned to the new treatment or a control within each trial subset (`subtrial' for short). 
Closed-form sample size formulae are derived to ensure each subtrial has a specified chance of correctly deciding whether the new treatment is superior to or not better than the control by some clinically relevant difference.
Given pre-specified levels of pairwise (in)commensurability, the subtrial sample sizes are solved simultaneously. 
The proposed Bayesian approach resembles the frequentist formulation of the problem in yielding comparable sample sizes for circumstances of no borrowing. 
When borrowing is enabled between commensurate subtrials, a considerably smaller trial sample size is required compared to the widely implemented approach of no borrowing.
We illustrate the use of our sample size formulae with two examples based on real basket trials. A comprehensive simulation study further shows that
the proposed methodology can maintain the true positive and false positive rates at desired levels.
\end{abstract}

\hspace{-1em} {\em Key words}:Borrowing strength; Commensurate priors; Master protocol; Mixture prior; Phase II.

\section{Introduction}
\label{sec:Intro}

Clinical research in precision medicine \citep{NEJM:Mirnezami2012, Nature:Schork2015} continues to thrive as a consequence of the rapid technological advances for identifying possible prognostic and predictive disease factors at the genetic level \citep{Nat:Aronson2015, Morganti2019}. Because of this,  an increasing number of biomarker-driven therapies have been formulated. In oncology, for example, much attention has been paid to therapies targeting one or multiple genomic aberrations \citep{BATTLEtrial2011, Nat:Hyman2018, LungMAP2020}. In contrast to conventional chemotherapy devised for treating histology-defined populations, such targeted therapies can potentially be beneficial to patients of various cancer (sub)types. 
Immune-mediated inflammatory diseases (IMIDs) \citep{NatRev:McInnes2021}  are another area where targeted therapies can be useful \citep{BMC:Grayling2021}. IMIDs generally involve a clinically diverse group of conditions that share common underlying pathogenetic features, calling for the development of effective immune-targeted therapeutics \citep{Nat:Pitzalis2020}. 
This paradigm shift towards precision medicine has challenged the use of traditional one-size-fits-all approaches to trial design, which aim to estimate the population average treatment effect. \\

Master protocols \citep{NEJM:Woodcock2017} comprise a class of innovative trial designs that address multiple investigational hypotheses.
Newly emerging types include {\em basket trials} that can simultaneously evaluate a new treatment in stratified patient subgroups displaying a common disease trait \citep{AO:Renfro2017, ARM:Tao2018}. An implication of the stratification is that patients may respond very differently to the same treatment due to their distinct disease subtypes, stages or status. Fully acknowledging the heterogeneity could lead to the use of stand-alone analyses that regard the stratified subgroups in isolation. Such an analysis strategy, though adopted in early basket trials, may not be ideal for realising the promise of basket trials. This is mainly because it (i) fails to treat the combined (sub)trial components as a single study, and (ii) often yields low-powered tests of the treatment effect, due to the small sample sizes.  \\

Sophisticated analysis models, which feature {\em borrowing of information} between subgroups (ideally between those with commensurate treatment effects only), have been proposed in the statistical literature. One pivotal strategy is to fit a Bayesian hierarchical random-effects model \citep{SiM:Thall2003, CT:Berry2013}, assuming that the subgroup-specific treatment effects are exchangeable, i.e., as random samples drawn from a common normal distribution with unknown mean and variance.
This methodology has been extended to involve (i) a finite mixture of exchangeability distributions, see, e.g., \citet{CCT:Liu2017, JRSSC:Chu2018, SiM:Jin2020}; as well as (ii) non-exchangeability distributions such that subgroups with an extreme treatment effect can be inferred to their own devices \citep{PS:Neuenschwander2016}. 
The former reflects the concern that some subsets of trial may be more commensurate between themselves than others.
A highly relevant proposal is to further cluster the subgroups, so that the corresponding treatment effects are assumed to be  exchangeable within the same cluster.
\citet{SMMR:Chen2020} present a two-step procedure, by which subgroups are clustered using a Bayesian nonparametric model, before fitting an adjusted Bayesian hierarchical random-effects model. \\

A few authors have further recommended to discuss the exchangeability or commensurability of any two or more subgroups. These proposals are for better characterisation of the complex trial data structure, which could involve mixtures of exchangeable or non-exchangeable patient subgroups.
\citet{BIOS:Psioda2019} apply a Bayesian model averaging technique \citep{SS:Hoeting1999} to accommodate the possibility that any configuration of subgroups may have the same or disparate response rate.  
\citet{SiM:Hobbs2018} construct a matrix containing elements with value of 0 or 1, indicating that any pair of subgroups can be exchangeable or non-exchangeable. 
Alternatively, \citet{BIOS:ZhengWason2019} propose measuring the pairwise (in)commensurability by distributional discrepancy to enable an appropriate
degree of borrowing from each complementary subgroup, which yields a largest weight allocated to the most commensurate one(s). \\

Development of methods to choose an appropriate sample size for basket trials, however, appears to fall behind. A widely implemented approach is to sum up the sample sizes, calculated as if these trial subsets are to be carried out as separate studies. 
Whilst this could impair the efficiency of decision making, alternative approaches to sample size determination that permit borrowing of information are lacking.
In this paper, we propose formal sample size planning for the design of basket trials. It strikes a balance between the sample size saving and the need of enrolling a sufficient number of patients to assure inferences about the subgroup-specific treatment effects.
As the importance of randomised controlled trials has been increasingly emphasised in oncology \citep{EJC:Ratain2009, JNCI:Grayling2019}, IMIDs \citep{BMC:Grayling2021} and rare-disease \citep{EJC:Prasad2015} research, this paper will focus on randomised basket trial designs with a primary objective of simultaneously comparing the new treatment against a control in various patient subgroups.
We will thus develop our sample size formulae, presuming that the analysis is performed using a model adapted from \citet{BIOS:ZhengWason2019}. \\

The remainder of this paper is organised as follows. In Section \ref{sec:methods}, we introduce a Bayesian model that estimates the treatment effect specific to subgroups using entire trial data, as well as the derivation of sample size formulae appropriate for basket trials. Two data examples are presented in Section \ref{sec:appli} to illustrate the use of our formulae for the design of randomised basket trials. In Section \ref{sec:sims}, we describe a simulation study that evaluates the operating characteristics of randomised basket trials. Finally, we conclude with a brief discussion and highlight several areas that deserve future research in Section \ref{sec:discuss}.

\section{Methods}
\label{sec:methods}

\subsection{Leveraging complementary subtrial data into commensurate priors}
\label{sec:cpp}

Let us consider the design of a basket trial where patients can be classified into $K$ subgroups.
These patients nonetheless share a common feature (e.g., a genetic aberration, clinical symptom or mechanism of drug action), on which a new targeted therapy may potentially improve patient outcomes.
Each component study in a distinct patient subgroup will hereafter be referred to as a trial subset (i.e., `subtrial' for short). Within each subtrial $k$, patients are randomised to receive either the experimental treatment (labelled $E$) with probability $R_k\in (0, 1)$, or a control (labelled $C$) with probability $(1 - R_k)$, for $k=1,\dots, K$.
We further assume that the measured responses are normally distributed with their own subtrial-specific parameters: $X_{ijk} \sim N(\mu_{jk}, \sigma_{k}^2)$, for $j = E, C; k = 1,\dots,K$.
Letting $n_{k}$ denote the subtrial sample size, the difference in mean responses is $\bar{X}_{Ek}-\bar{X}_{Ck}\sim N\left(\mu_{Ek}-\mu_{Ck}, \frac{\sigma_{k}^2}{n_kR_k(1-R_k)}\right)$, per subtrial $k$.
For the ease of notation, we let $\theta_k = \mu_{Ek} - \mu_{Ck}$ denote the treatment effect for subtrial $k$.  
It is important to clarify at the outset that this design aims to estimate the subtrial-specific treatment effects, i.e., $\theta_1, \dots, \theta_K$, instead of an overall treatment effect averaged over all subtrials.
If permitting borrowing of information across subtrials, these treatment effects are to be estimated using the entire trial data (with $\sum_{k=1}^K n_k$ patients) rather than in isolation (with $n_1, \dots, n_K$ patients, respectively). \\

We follow \citet{BIOS:ZhengWason2019} in specifying commensurate priors for each $\theta_k$, using information from the $(K-1)$ complementary subtrials indexed by $q\neq k, \, \forall k = 1, \dots, K$. 
This methodology regards any $\theta_q$ as a biased representation of $\theta_k$, yet the direction and the size of such bias are unknown \citep{BIOM:Hobbs2011}. 
More specifically, these commensurate priors are formulated as conditional normal distributions that are centred at $\theta_q$s, respectively; whilst the precisions (i.e., reciprocal of variances), denoted by $\nu_{qk}$, accommodate the heterogeneity between two subtrials $k$ and $q$.  
Our commensurate prior models for the continuous location parameter $\theta_k$ can thus be given by
\begin{equation}
\begin{split}
\theta_{k} \mid \theta_q, \nu_{qk} & \sim N(\theta_q, \nu_{qk}^{-1}), \qquad \forall k = 1, \dots, K, \\
\nu_{qk} & \sim w_{qk} \text{Gamma}(a_{1}, b_{1}) + (1-w_{qk})\text{Gamma}(a_{2}, b_{2}),
\end{split}
\label{eq:commenprio}
\end{equation}

\noindent where a two-component Gamma mixture prior (with $a_1/b_1$ and $a_2/b_2$ being the respective means of the component distributions), instead of a spike-and-slab prior in the original proposal, is placed on each $\nu_{qk}$ for the convenience of analytic tractability \citep{JASA:Zheng2020}.
In particular, these two Gamma mixture components correspond to extreme cases of substantial or limited discounting of information from a complementary subtrial $q$. 
For illustration, we suppose that the first Gamma mixture component 
has its density massively on small values, and the second component
on large values.
The prior mixture weight $w_{qk}\in [0, 1]$, which plays a role of balancing between the extreme cases, can thus reflect one's preliminary scepticism about the degree of commensurability between $\theta_k$ and $\theta_q$.
That is, when subtrials $k$ and $q$ are thought of as incommensurate (commensurate), $w_{qk}$ can be set close to 1 (0), thus forcing the conditional prior variance $\nu_{qk}^{-1}$ towards large (small) values for substantial (limited) discounting. \\

By integrating out $\nu_{qk}$, the conditional prior for $\theta_{k}$ given $\theta_q$ only follows a shifted and scaled $t$ mixture distribution, with its two components both centred at $\theta_q$. This unimodal $t$ mixture distribution can further be approximated by matching the first two moments to give
\begin{equation}
\theta_k \mid \theta_q  \, \dot\sim  \, N\left(\theta_q, \frac{w_{qk}b_{1}}{a_{1}-1} +\frac{(1-w_{qk})b_{2}}{a_{2}-1} \right), \quad \text{ with } a_{1}, a_{2} > 1,
\label{eq:margmui}
\end{equation}

\noindent which incorporates the respective variances of the $t$ component distributions. As has been shown by \citet{JASA:Zheng2020}, this normal approximation provides good properties for the coverage of credible intervals of interest.
Note that the location of each commensurate prior, $\theta_q$, is an unknown parameter. It captures the information from a complementary subtrial $q$, of which the required sample size $n_q$ as well as the allocation proportion $R_q$ is yet to be determined. \\

Let $\boldsymbol{x}_q = \{x_{1Eq}, \dots, x_{n_qEq}; x_{1Cq}, \dots, x_{n_qCq} \}$ denote the data of a complementary subtrial $q$. 
We consider the difference of sample means, $\bar{X}_{qE}-\bar{X}_{qC}$, as the random variable to draw the Bayesian inference. 
With an `uninformative' operational prior $\theta_q\sim N(m_{0q}, s_{0q}^2)$, we derive the posterior as
\begin{equation}
\begin{split}
\theta_q \mid \boldsymbol{x}_q &\sim N\left(\lambda_q, \left(\frac{1}{s_{0q}^2} + \frac{n_qR_q(1-R_q)}{\sigma_{q}^2}\right)^{-1} \right), \\
\text{with  } \qquad \lambda_q &= \frac{m_{0q}}{1+ \frac{s_{0q}^2}{\sigma_{q}^2}\cdot n_qR_q(1-R_q)}+\frac{\bar{x}_{Eq} - \bar{x}_{Cq}}{1+\frac{\sigma_{q}^2}{s_{0q}^2}\cdot\left[n_qR_q(1-R_q)\right]^{-1}},
\end{split}
\label{eq:oppost}
\end{equation}

\noindent where $\bar{x}_{jq}$ denotes the average response of samples by treatment group $j = E, C,$ within subtrial $q$.
Combining \eqref{eq:margmui} and \eqref{eq:oppost}, we obtain a commensurate prior for $\theta_k$ that leverages data of a complementary subtrial $q\neq k$:
\begin{equation}
\begin{split}
\theta_k \mid \boldsymbol{x}_q &\sim N(\lambda_q, \xi_{qk}^2), \qquad  \forall k = 1, \dots, K, \\
\text{with } \qquad \xi_{qk}^2 &= \left(\frac{1}{s_{0q}^2} + \frac{n_qR_q(1-R_q)}{\sigma_{q}^2} \right)^{-1} + \frac{w_{qk}b_{1}}{a_{1}-1} +\frac{(1-w_{qk})b_{2}}{a_{2}-1}.
\end{split}
\label{eq:borrowprior}
\end{equation}

Consider now borrowing information from all complementary subtrials, with $K \geq 3$. Let $\boldsymbol{x}_{(-k)}$ denote the data from all subtrials excluding $k$, that is, all the $(K-1)$ sets of complementary data for subtrial $k$. By the convolution operator \citep{Grinstead1997}, we stipulate a collective, commensurate prior for leveraging all complementary subtrial data:
\begin{equation}
\theta_k \mid \boldsymbol{x}_{(-k)} \sim N\left(\sum_{q\neq k} p_{qk} \lambda_q, \sum_{q\neq k} p_{qk}^2 \xi_{qk}^2 \right), \qquad  \forall k = 1, \dots, K,
\label{eq:collprior}
\end{equation}

\noindent where $p_{qk}$ are the synthesis weights, with $\sum_q p_{qk} = 1$, assigned to the respective commensurate priors specified using $\boldsymbol{x}_q$. 
These synthesis weights can be transformed from the chosen values for $w_{qk}$ in the commensurate prior models, following an objective-directed approach \citep{BIOS:ZhengWason2019}.
More specifically, we expect the largest synthesis weight, $p_{qk}$, is assigned to the most commensurate prior $N(\lambda_q, \xi_{qk}^2)$, specified based on a subtrial $q\neq k$ that manifests the smallest discrepancy with subtrial $k$ out of all the $(K-1)$ complementary subtrials.
Recall each $w_{qk}$, as one key parameter to determine $N(\lambda_q, \xi_{qk}^2)$, would have been chosen to appropriately reflect the pairwise discrepancy (i.e., incommensurability). One may apply a decreasing function of $w_{qk}$ to compute $p_{qk}$.
A $K\times K$ matrix can be constructed to contain all $w_{qk}$ in column $k$ and row $q$ as:
$$
\begin{pmatrix}
0 & w_{12} & \cdots & w_{1K} \\
w_{21} & 0 & \cdots & w_{2K} \\
\vdots  & \vdots  & \ddots & \vdots  \\
w_{K1} & w_{K2} & \cdots & 0
\end{pmatrix}.
$$
We note that this matrix should be symmetric with $w_{qk} = w_{kq}$, since each is intended to reflect the level of pairwise incommensurability. That is, the magnitude of incommensurability between subtrials $k$ and $q$ is the same as that between $q$ and $k$. If stratifying the matrix by column,
the off-diagonal elements in column $k$ represent the postulated levels of discounting with respect to the complementary subtrial data. Recall that the latter have been used to specify the respective commensurate priors in the form of \eqref{eq:borrowprior}, with $q\neq k$.
The decreasing function given by 
$$
p_{qk} = \frac{\exp(-w_{qk}^2/c_0)}{\sum_q \exp(-w_{qk}^2/c_0)}, \qquad  \forall k = 1, \dots, K,
$$
has been illustrated to have satisfactory properties \citep{BIOS:ZhengWason2019, JASA:Zheng2020}.
The concentration parameter $c_0$, if set equal to a value close to $0^+$, appropriately discerns the $(K-1)$ values of $w_{qk}$; thus, a $p_{qk}\rightarrow 1$ would be assigned to the corresponding commensurate prior for $\theta_k$ based on $\boldsymbol{x}_q$, in which the smallest $w_{qk}$ has been used. Otherwise, a value of $c_0\gg w_{qk}$ yields nearly all $p_{qk}$ to equal $1/(K-1)$ irrespective of the values of $w_{qk}$. Moreover, this transformation yields equal $p_{qk}$ when all $w_{qk}$ are equal. We generally recommend setting $c_0$ to a value that is substantially smaller than the magnitude of $w_{qk}$; for a thorough evaluation of performance by varying $c_0$, we refer the reader to \citet{BIOS:ZhengWason2019}. \\

By using the Bayes' Theorem, the collective commensurate prior in the form of \eqref{eq:collprior} will be updated by the contemporary subtrial data $\boldsymbol{x}_k$ to give the posterior distribution as
\begin{equation}
\theta_k \mid \boldsymbol{x}_k, \boldsymbol{x}_{(-k)} \sim N \left(d_{\theta_k}, 
\left(\frac{1}{\sum_q p_{qk}^2 \xi_{qk}^2} + \frac{n_kR_k(1-R_k)}{\sigma_{k}^2} \right)^{-1}
\right).
\label{eq:collpost}
\end{equation}
\noindent The posterior mean is a convex combination of the prior mean $\sum p_{qk}\lambda_q\, (q \neq k)$  and the data likelihood. We will give the exact expression of $d_{\theta_k}$ in Section \ref{sec:sims} to carry out the decision making for simulated trials.

\subsection{Sample size formulae for basket trials comparing two normal means}
\label{sec:ssd}

The frequentist approach to sample size determination makes use of hypothesis testing, with $H_{0k}: \theta_k \leq 0$ against $H_{1k}: \theta_k > 0$, if assuming that greater values of $X_{ijk}$ indicate better effect. 
In this traditional framework, a study sample size is computed such that the treatment effect, $\theta_k$, will be found significant at a level $\alpha$ with probability $1-\beta$, given a certain magnitude of the treatment effect considered clinically meaningful. \\

We follow the Bayesian decision making framework, presented by \citet{SiM:Whitehead2008}, to compute two interval probabilities from our posterior distribution as derived in \eqref{eq:collpost}, so that the subtrial sample sizes, $n_1, \dots, n_K$, are sought for providing compelling evidence of $E$ being {\em either} superior to {\em or} not better than $C$ by some magnitude $\delta$ in each subtrial $k = 1, \dots, K$.  
The posterior distribution of $\theta_k$ specific to each subtrial $k=1, \dots,K$ will thus be evaluated to declare that $E$ is 
\begin{align}
\text{(i) efficacious, if} \quad & \mathbb{P}(\theta_k > 0 \mid \boldsymbol{x}_k, \boldsymbol{x}_{(-k)}) \geq \eta, \label{eq:criterion1} \\
\text{(ii) futile, if} \quad & \mathbb{P}(\theta_k \leq \delta \mid \boldsymbol{x}_k, \boldsymbol{x}_{(-k)}) \geq \zeta, \label{eq:criterion2} 
\end{align}
\noindent where $\eta$ and $\zeta$ are probability thresholds for the success and futility criteria, respectively. 
By using this decision rule, the two posterior tail probabilities of $\theta_k$ are controlled. Specifically, we desire the area under the density from $-\infty$ to the left of 0 to be limited below $1-\eta$, with the area under the density from $\infty$ to the right of $\delta$ to be below $1-\zeta$.
The sample size therefore needs to be sufficiently large for a decisive declaration of the treatment effectiveness or futility per subtrial $k$. That is, $d_{\theta_k}/\sigma_{\theta_k} \geq z_\eta$ or $(\delta - d_{\theta_k})/\sigma_{\theta_k} \geq z_\zeta$ should be guaranteed.
Here, $z_\eta$ satisfies $\Phi(z_\eta) = \eta$ where $\Phi(\cdot)$ denotes the standard normal distribution function, with $z_{\zeta}$ defined similarly.
We thus require $\delta/\sigma_{\theta_k} \geq z_\zeta + z_\eta$, which leads to 
\begin{equation}
\frac{1}{\sigma_{\theta_k}^2} \geq \left( \frac{z_\eta + z_\zeta}{\delta}\right)^2.
\label{eq:SSbase}
\end{equation}
The left-hand side of \eqref{eq:SSbase} is precisely the posterior precision for $\theta_k$. \\

When borrowing of information is not permitted, $\mathbb{P}(\theta_k >0 \mid \boldsymbol{x}_k)$ and $\mathbb{P}(\theta_k \leq \delta \mid \boldsymbol{x}_k)$ are computed instead. Thus, $\frac{1}{\sigma_{\theta_k}^2} = \frac{1}{s_{0k}^2} + \frac{n_kR_k(1-R_k)}{\sigma_{k}^2}$, which can be rearranged to give
\begin{equation}
n_k^{0} \geq \frac{\sigma_{k}^2}{R_k(1-R_k)}\left[\left(\frac{z_\eta + z_\zeta}{\delta}\right)^2 - \frac{1}{s_{0k}^2} \right], \quad \forall k = 1, \dots, K.
\label{NoBrwSSD}
\end{equation}
By contrast, based on the proposed Bayesian model for borrowing of information, $\sigma_{\theta_k}^2$ comes from the closed-form expression in \eqref{eq:collpost}.
This leads to
\begin{equation}
n_k \geq \frac{\sigma_{k}^2}{R_k(1-R_k)}\left[\left(\frac{z_\eta + z_\zeta}{\delta}\right)^2 - \frac{1}{\sum_q p_{qk}^2 \xi_{qk}^2} \right], \quad \forall k = 1, \dots, K, \; q\neq k,
\label{eq:BayesSSD}
\end{equation}
which looks similar to \eqref{NoBrwSSD}, but involves the commensurate prior variances $\xi_{qk}^2$ in the form of \eqref{eq:borrowprior}. The latter leverages the complementary subtrial information. 
Thus, a smaller integer for $n_k$ could be expected if the complementary subtrials, labelled $q\neq k$, are to collect rich information and, further, considerable borrowing of information happens. 
To ensure the inference in all $K$ subtrials, we require that $\forall k = 1, \dots, K$,
\begin{equation*}
\frac{n_kR_k(1-R_k)}{\sigma_{k}^2} + \left[\sum_q p_{qk}^2 \left(\left(\frac{1}{s_{0q}^2} + \frac{n_qR_q(1-R_q)}{\sigma_{q}^2} \right)^{-1} + \frac{w_{qk}b_{1}}{a_{1}-1} +\frac{(1-w_{qk})b_{2}}{a_{2}-1} \right)\right]^{-1}  \geq \left(\frac{z_\eta + z_\zeta}{\delta}\right)^2, 
\end{equation*}
\noindent with $p_{qk}$ transformed from $w_{qk}$ following the stipulation in Section \ref{sec:cpp}. 
The $K$ nonlinear equations of $n_k$ and its $n_q$s are continuously differentiable. 
We apply Newton's method for systems of nonlinear equations \citep{NM:Dennis1983} to find $n_1, \dots, n_K$ that satisfy the $K$ constraints simultaneously, given known $w_{qk}$ and  $\sigma_k^2$. \\

The importance of $w_{qk}$, in computing subtrial sample sizes is of particular interest. 
Whilst the definition of data (in)commensurability can vary on a case-by-case basis, these values may better be specified in collaboration with a subject-matter expert. Those conversations may also help quantify the magnitude of incommensurability, particularly in the absence of pilot data or relevant investigation. We caution that the choice of $w_{qk}$ should reflect the level of pairwise (in)commensurability {\em a priori} in the practical implementation, rather than for the desire of obtaining a minimal sample size.

\section{Application}
\label{sec:appli}

\subsection{A UK-based basket trial for treating patients with chronic diseases}
\label{sec:OACS}

The randomised, placebo-controlled Obeticholic acid for the Amelioration of Cognitive Symptoms trial (known as the `OACS trial', ISRCTN15223158) aims to assess the efficacy of Obeticholic acid ($E$), as compared to a placebo ($C$), for treating cognitive deficits.
The OACS trial is split into three subtrials, each focusing on a distinct patient subpopulation defined by the disease stage and clinical area. Namely, OACS-1 for patients with stabilised primary biliary cholangitis (PBC) $>2$ years since diagnosis; OACS-2 for patients with new-onset PBC $\leq 2$ years; and OACS-3 for patients with Parkinson's disease.
The primary outcome is a composite cognitive test score obtained from the CANTAB platform \citep{Goldberg2013}, which is an extensively used tool in clinical practice. 
The reduction in the composite CANTAB score from the baseline, after 26 weeks of treatment, will be analysed as a normally distributed primary endpoint. 
We assume the magnitude of such reduction can be adequately depicted by values ranging from -5 to 5, where a high value suggests improvement of cognitive symptoms in a patient. \\

The sample sizes have been determined as 40 (20 on $E$ and $C$ each) for OACS-1, 25 (15 on $E$ and 10 on $C$) for OACS-2, 25 (15 on $E$ and 10 on $C$) for OACS-3, assuming that these subtrials are to be analysed on their own.
As specified in the trial protocol, these were not computed based on hypothesis testing considerations originally. However, the resulting sample sizes are consistent with $\sigma_1^2 = 6.177,\, \sigma_2^2 = 5.134,\, \sigma_3^2 = 5.134$ to ensure 90\% statistical power for OACS-1 and 80\% for the the other subtrials to detect a difference of $\delta = 2.3$, whilst controlling the type I error rate below 0.05. 
In the following, we use these quantities to illustrate the application of the proposed methodology, if the OACS basket trial would have been designed using the Bayesian decision making framework outlined above. \\

We set $s_{0k}^2 = 100$ and $\nu_{qk} \sim w_{qk} \text{Gamma}(1.1, 1.1) + (1-w_{qk})\text{Gamma}(54, 3)$ for all $k = 1, \dots, 3$. 
Specifically, the two Gamma mixture components correspond to the mean and a 95\% credible interval for each $\nu_{qk}$ as 1 [0.041, 4.286] and 18 [13.522, 23.108], respectively. This accommodates two extreme cases for limited or substantial borrowing when $w_{qk} = 1$ or 0.\\ 

If no borrowing is permitted, $n_k^0 = 39.8,\, 24.8,\, 24.8$ according to \eqref{NoBrwSSD}. This is about the same as the actual sample sizes of 40, 25, 25, respectively. For illustration purpose only, we use the following matrix of $w_{qk}$ 
$$
\begin{pmatrix}
w_{11} & w_{12} & w_{13} \\
w_{21} & w_{22} & w_{23} \\
w_{31} & w_{32} & w_{33} \\
\end{pmatrix} = 
\begin{pmatrix}
0       & 0.239     & 0.417     \\
0.239     & 0     & 0.145   \\
0.417 & 0.145 & 0	  
\end{pmatrix},
$$
and $c_0 = 0.05$ to compute the synthesis weights $p_{qk}$, following the objective-directed approach outlined in Section \ref{sec:cpp}. The
subtrial sample sizes are found to be $n_k = 33.3, \, 11.8,\, 18.2$, fixing $\eta = 0.95$ and $\zeta = 0.90$ for subtrial 1 or 0.80 for subtrial 2 maintaining the same treatment allocation ratios $R_k = 0.5, \, 0.6, \, 0.6$.
These are considerably smaller than the subtrial sample sizes assuming no borrowing or the frequentist counterparts.

\subsection{Simultaneous evaluation of a new inhibitor in seven cancer subtypes}
\label{sec:summit}

The ongoing SUMMIT basket trial (NCT01953926) adopted a single-arm design to evaluate a new pan-HER kinase inhibitor neratinib in 141 patients with HER2-mutant or HER3-mutant tumours \citep{Nat:Hyman2018}.
A binary outcome (i.e., responder or no responder, corresponding to a tumour shrinkage $\geq$ 30\% or below) was used in line with the RECIST criteria \citep{EJC:Eisenhauer2009}.
The SUMMIT trial additionally reported analysis on secondary outcomes, which include the change in tumour volume on a continuous scale of -100\% to 100\%.
We assume a new randomised basket trial would follow, wherein the change in tumour volume from -100\% to 100\% is the primary outcome. A negative sign indicates the clinical benefit, since it is hoped that the tumour shrinks from the baseline measurement due to the treatment. 
With $\delta < 0$, the trial decision criterion expressed by \eqref{eq:criterion1} and \eqref{eq:criterion2} should be altered as (i) efficacious, if $\mathbb{P}(\theta_k < 0 \mid \boldsymbol{x}_k, \boldsymbol{x}_{(-k)}) \geq \eta$, and (ii) futile, if $\mathbb{P}(\theta_k \geq \delta \mid \boldsymbol{x}_k, \boldsymbol{x}_{(-k)}) \geq \zeta$.\\

We narrow the focus on seven of the originally investigated 21 cancer subtypes only, of which the mean responses for patients receiving neratinib ($E$) were approximately $\mu_{Ek} = -0.489, 0.226$, $-0.181, 0.293$, $0.329, -0.275, -0.136$. 
We further assume that the mean responses on a control treatment embedded in the newly planned basket trial are $\mu_{Ck} = 0$, and that patients within each subtrial have equal probability to receive $E$ or $C$; that is, $R_k = 0.5$ for $k = 1, \dots, 7$. \\

Based on the published SUMMIT trial results \citep{Nat:Hyman2018}, we assume the subtrial-specific variances are $\sigma_{k}^2 = 0.587^2$, $0.345^2,\,  0.380^2,\, 0.347^2,\, 0.344^2,\, 0.392^2,\,  0.392^2$.
The pairwise discrepancy between the assumed outcome distributions, $N(\mu_{Ek}, \sigma_{k}^2)$, can be quantified using, e.g., the Hellinger distance \citep{Hellinger:Dey1994},
$$
w_{qk} = \left[1 - \sqrt{\frac{2\sigma_q \sigma_k}{\sigma_q^2 + \sigma_k^2}}\exp \left( -\frac{(\mu_{Eq} - \mu_{Ek})^2}{4(\sigma_q^2 + \sigma_k^2)} \right) \right]^{1/2}.
$$
By targeting $\delta = -0.40$ and retaining the specification of other parameters from Section \ref{sec:OACS}, $n_k = 52.0, \, 17.3, \, 20.5, \, 17.0, \, 17.1, \, 22.5, \, 22.0$. 
If no borrowing is permitted, $n_k^0 = 53.3, \, 18.4, \, 22.3, \, 18.6$, $18.3, \, 23.8, \, 23.8$.
Only a small reduction in the sample sizes has been observed, largely because most of the values of $w_{qk}$ are above 0.30.

\section{Simulation study}
\label{sec:sims}

\subsection{Basic setting}

Motivated by the SUMMIT trial, we consider the sample size planning of  basket trials following the same data structure.
That is, the basket trial would enroll $n_1, \dots, n_7$ patients to the respective subtrials, with $R_k = 0.5$ in all $K = 7$ subgroups, under six possible scenarios.
Figure \ref{fig:SimSc} visualises the six simulation scenarios, where the location and length of lines suggest the distributions of $X_{ijk}, \, j = E, C$, while a larger bubble corresponds to a larger value of $w_{qk}$.
Here, we have followed the specification of $w_{qk}$ given in Section \ref{sec:summit} to compute the pairwise Hellinger distance to characterise (in)commensurablity and obtain $w_{qk}$ for the levels of borrowing/discounting strength. 
Scenarios 4 and 6 correspond to two special cases of the treatment being consistently effective (alternative hypotheses) and consistently futile (null hypotheses), respectively.
Both scenarios feature perfect commensurability; that is, the outcomes $X_{iEk}$ and $X_{iCk}$ have their respective, same distribution across subtrials, so all $w_{qk} = 0$. Scenario 5 represents a mixed null situation, where $\theta_k = 0$ holds for four of the subtrials only. The other scenarios are constructed to reflect various levels of data incommensurability.
Exact configurations of these simulation scenarios, i.e, values of $\mu_{Ek}$ along with all $\mu_{Ck} = 0$ as well as the subtrial-specific variances $\sigma_k^2$, are listed in Table S1 of the Supplementary Materials. \\

[Figure \ref{fig:SimSc} about here.] \\

We retain the prior specification and the probability thresholds unchanged from Section \ref{sec:appli}. Table \ref{tab:SimSS} thus gives the subtrial sample sizes required to reach a decisive conclusion of either $E$ is superior to or not better than $C$ by $\delta = -0.4$ using the respective sample size formulae.
Because no $w_{qk}$ has been set to 1 in any scenario, the sample sizes $n_k$ computed from the approach of borrowing are generally smaller than $n_k^0$ from the approach of no borrowing. 
Scenario 1 was constructed from the illustrative data example in Section \ref{sec:summit}. Since relatively large values have been chosen for $w_{qk}$, only a slight decrease in sample sizes is observed. Unlike  Scenario 1 that displays divergent effects, Scenario 2 is featured with a higher degree of commensurability that $E$ has an enhanced benefit over $C$ in all subtrials. A smaller trial sample size is then required.
Scenario 3 has all the variances $\sigma_k^2 = 0.3$. Consequently, $n_k^0$, based on the approach of no borrowing, are solved to be equal to 46.4 for all subtrials. Whereas, using the proposed methodology, the sample sizes for subtrials 1, 3, 4 and 6 are smaller, as these are recognised to be more commensurate between themselves than with the other three. A similar explanation can be given to Scenario 5: subtrials 2, 4 and 5 have greater sample size savings because the corresponding $w_{qk}$ takes smaller values.
Scenarios 4 and 6 represent the situations of perfect commensurability. With all $w_{qk} = 0$, a substantial reduction in the subtrial sample sizes is observed. \\

[Table \ref{tab:SimSS} about here.] \\

In the numerical evaluation below, we simulate the outcomes $X_{ijk}, j = E, C,$ from $N(\mu_{Ek}, \sigma_k^2)$ and $N(\mu_{Ck}, \sigma_k^2)$ for patient $i = 1, \dots, n_k,$ within subtrial $k = 1, \dots, 7$.
For each scenario, 100,000 replicates of the basket trials are simulated to fit:
\begin{itemize}
\item the proposed Bayesian model, which yields the posterior distributions for $\theta_k$ in the form of \eqref{eq:collpost}, with 
$$
d_{\theta_k} = \frac{\sigma_k^2/(n_kR_k(1-R_k))\cdot \sum_q p_{qk}\lambda_q + (\bar{x}_{Ek} - \bar{x}_{Ck}) \cdot \sum_q p_{qk}^2 \xi_{qk}^2}{\sum_q p_{qk}^2 \xi_{qk}^2 + \sigma_k^2/(n_kR_k(1-R_k))},
$$
\item a Bayesian stand-alone analysis model for no borrowing of information. Operational priors, i.e., $N(m_{0k}, s_{0k}^2)$, are placed on each $\theta_k$. This leads to the posterior distributions for $\theta_k$ based on $\boldsymbol{x}_k$ alone, which has the same form as \eqref{eq:oppost}, with the subscript $q$ replaced by $k$. We set all $m_{0k} = 0$ in the simulations.
\end{itemize}

\subsection{Results}

We summarise the frequency of simulated trials concluding that $E$ is either efficacious or futile, based on the 100,000 replicates per scenario and model. 
Figure \ref{fig:SimRes} depicts the percentages of (sub)trials declaring effectiveness of $E$ and those declaring futility. Wherever the lengths of bars sum up to 100\%, this means the study is planned with an adequate sample size for decisive decision making.
As we can observe, collecting data from $n_1, \dots, n_7$ patients to fit the proposed analysis model ensures 100\% of the (sub)trials to conclude that $E$ is either superior to or not better than $C$ by $\delta$. 
Whereas, it is not the case (i.e., all below 100\%) if implementing the Bayesian model for no borrowing, since larger sample sizes (i.e., $n_k^0$ in Table \ref{tab:SimSS}) would be required to ensure the same level of posterior distribution informativeness for the trial decision.
In Scenarios 1 and 2 where $n_k$ and $n_k^0$ were comparable, these two Bayesian models yielded comparable proportions of (sub)trials with a decisive trial decision.
Yet in Scenarios 4 and 6 where substantial sample size saving were made, a disparity is observed, because the posterior distributions for $\theta_k$ based on $\boldsymbol{x}_k$ alone are far less informative than those based on $\boldsymbol{x}_1, \dots, \boldsymbol{x}_K$. \\

[Figure \ref{fig:SimRes} about here.] \\

In Scenarios 2 and 3, $E$ is potentially superior than $C$ yet the magnitude tends to be smaller than desired on average.
Only subtrial 1 has a mean treatment effect greater than $\delta$, so about 91.2\% of the simulated (sub)trials have declared $E$ being effective. By contrast, subtrials 2 and 5 have mean treatment effects closest to 0 and $\delta$, respectively. Therefore, subtrial 2 has higher chance to declare futility than effectiveness of $E$, but subtrial 5 is on the contrary.
Scenario 4  mimics the borderline case where the mean treatment effect just has the size of $\delta$. Using the proposed methodology, about 82.1\% of the simulated (sub)trials have favoured $E$ for effectiveness in all seven subtrials. 
These subtrialwise true positive rates are about equal to our chosen threshold $\zeta = 0.80$. 
Scenario 5 assumes a mixture of subtrial-specific treatment effects with $\theta_k = 0$ or $\geq \delta$.
Referring to subtrials 2, 4, 5 and 7, less than 5\% of the simulated trials conclude effectiveness erroneously. 
The two Bayesian models yield similar operating characteristics in this scenario, as the computed $n_k$ and $n_k^0$ were close.
In Scenario 6, the proportion of incorrect decision of effectiveness is maintained below 5\% for all subtrials using the proposed methodology. 
Unsurprisingly, using the approach of no borrowing to analyse the basket trial from only 62.3 patients has much lower chance of obtaining a definitive conclusion. 
The overall false positive rate (i.e., probability of incorrectly rejecting at least one subtrial that has true $\theta_k = 0$), based on the proposed methodology, is 0.150 for Scenario 5 and 0.054 for Scenario 6.  These increase to 0.192 and 0.346, respectively, if the approach of no borrowing is implemented instead. This is not surprising because the sample sizes were computed to control the error rate at the subtrial level. For strong control of the overall error rate, multiplicity adjustment such as the Bonferroni procedure is required. After the correction, one can still expect more benefit from borrowing information than not.\\

Focusing on Scenarios 4 -- 6 for the true positive and false positive rates at the subtrial levels, the proportions are not exactly 80\% or 5\% because of the simulation randomness. Additional simulations have been carried out for homoscedastic scenarios by varying the value of $\sigma_k^2$. 
Figure \ref{fig:TPFP} shows (i) the subtrialwise sample sizes, $n_k$, determined based on our sample size formula in \eqref{eq:BayesSSD}, and correspondingly, (ii) the subtrialwise true positive and false positive rates based on the simulated 100,000 replicates of basket trials. Each set of the additional simulations yielded seven points (for $K=7$ subtrials), which congregate at the levels around $\zeta = 0.80$ or $1-\eta = 0.05$. In summary, the proposed methodology can lead to the control of error rates. \\

[Figure \ref{fig:TPFP} about here.] \\

We have also performed a sensitivity analysis to understand the effect of misspecified values of $w_{qk}$.
Table S2 in the Supplementary Materials reveals that the proposed methodology is reasonably robust to the misspecification of $w_{qk}$. Nonetheless, care is needed when the value of $w_{qk}$ in the analysis deviates too far from that used in the design.
When $w_{qk}$ is set to a larger value in the analysis than in the design (i.e., less borrowing is implemented than planned), a smaller percentage of trials conclude with a decisive decision. Whereas, a smaller value of $w_{qk}$ would yield a more informative posterior distribution, but this sometimes produces ambiguous conclusion of effectiveness or futility.

\section{Discussion}
\label{sec:discuss}

The importance of choosing an appropriate sample size can never be overemphasised \citep{Senn2007}. 
Whilst basket trials have major infrastructural and logistical advantages, sophisticated statistical models are needed for the sample size planning to preserve the added efficiency.
The most widely-used approach to date is based on a Bayesian stand-alone analysis model, which does not support information sharing across subtrials with commensurate treatment effects.
Consequently, the majority of basket trials recruit a higher sample size than required.
This not only causes a waste of resources, but could sometimes be unethical for exposing more patients than is necessary to a treatment that is yet to be fully approved \citep{BMJ:Altman1980}. 
To realise the promise of basket trials, this paper establishes a closed-form solution to the simultaneous determination of subtrial sample sizes.
The simulation study shows that the proposed methodology requires a smaller trial sample size wherever $0\leq w_{qk}<1$, without undermining the chance of detecting if there exists a clinically relevant difference between the experimental treatment and the control.  \\

For deriving our sample size formulae, we adopted the Bayesian decision making scheme elaborated by \citet{SiM:Whitehead2008}. Specifically, it involves two probability thresholds $\eta$ and $\zeta$ for reaching a decisive statement on the treatment's effectiveness or futility. 
In our numerical illustration, we set $\eta = 0.95$ and $\zeta = 0.80$ because these probability quantities yield $n_k^0$, obtained based on the approach of no borrowing, comparable to the frequentist solution of sample sizes with $\alpha = 0.05$ and $\beta = 0.20$.
Other choices may certainly be feasible: there is no conventional level to set these probability thresholds.
In practice, these quantities might be difficult to justify: fixing $\eta$ to, for example, 0.95 might mean a considerable increase in sample size compared to 0.90.
Since the sample sizes also depend on other parameters, we recommend the user generates plots for their cases following the pattern of our Figure S2 in the Supplementary Materials.\\

Two sets of key parameters to implement the proposed methodology are the variances $\sigma_k^2$ and the levels of pairwise discounting $w_{qk}$. 
Similar to the widely-used frequentist formulae \citep{Chow2007}, an increase in $\sigma_k^2$ would mean that larger sample sizes are needed to maintain the same level of precision in data.
We have restricted our focus on known variances throughout, since this is common in conducting clinical trials for most circumstances. 
Appropriate values for $\sigma_k^2$ to compute the subtrial sample sizes can be informed by pilot data or information from relevant investigations.
If retaining $\sigma_k^2$ as unknown parameters, priors about their magnitude would be required. 
Subtrial sample sizes would then be sought by controlling the average property of posterior interval probabilities of $\theta_k$ with respect to 0 and $\delta$, since these nuisance parameters need to be integrated out from the posterior.  \citet{JASA:Zheng2020} derived sample size formulae with unknown $\sigma_k^2$ for a relevant context.
Although different decision criteria were considered, one could follow their methodology to obtain the marginal posterior distributions, for which external information on $\sigma_k^2$  may further be incorporated.
For wider applications, we have extended the proposed methodology for basket trials using a binary (in both the randomised controlled and single-arm settings) or time-to-event outcome; see Sections C -- E of the Supplementary Materials for the corresponding sample size formulae. In the meanwhile, we note there are limitations; for example, the censoring assumptions for time-to-event data are greatly simplified. We hope this work will stimulate further research within this Bayesian decesion framework. \\

In the present work, data are supposed to be analysed after the completion of all subtrials. However, in practice, certain subtrials may take much longer to complete recruitment due to low prevalence.
One could (a) adopt a `first (subtrials) complete, first analysed' principle, or (b) alter the constraint for simultaneously solving $n_1, \dots, n_K$, e.g., by making them proportional to the prevalence of respective target subpopulations, whilst maintaining an overall statistical power or decision accuracy.
For strategy (a), more borrowing would be possible from subtrials that complete faster to those that complete slower. With strategy (b), all subtrials may finish about the same time to yield a joint data analysis. 
We should note that it is not obvious if either strategy leads to a substantial increase in the total sample size.  \\

When borrowing of information is permitted, a reduced sample size can be expected by setting $w_{qk} < 1$. The smaller the values of $w_{qk}$, the more borrowing is possible. The present methodology requires that these values  be specified to reflect the pairwise (in)commensurability of subtrial data. This is especially feasible when a pilot study has been conducted. More details about the practical implementation, particularly the specification of parameters, are available in Section F of the Supplementary Materials.
Throughout, we have elaborated the methodology concerning a pre-specified magnitude  relating to the effect size, $\delta$, to find subtrial sample sizes. 
Extending the calculation to consider subtrial-specific effective sizes, say, $\delta_k$, is straightforward. A smaller value of $\delta_k$ would indicate that a larger $n_k$ is needed, if all other parameters are held fixed. For practical implementation, the user may substitute the corresponding argument (currently as a single value) by a vector in the openly available software. \\

Our sensitivity analysis in Section G of the Supplementary Materials suggests the proposed methodology is reasonably robust against misspecification of $w_{qk}$. 
Nevertheless, when the values deviate too far from the truth, the resulting sample sizes would not reflect what is needed to achive the trial's objectives. 
One avenue for future research would therefore be developing methodology for sample size reassessment in basket trials.
Practitioners may start with rather conservative choices of $w_{qk}$ assuming limited borrowing, and re-estimate $w_{qk}$ using accumulating data from the ongoing trial at interims. 
As the reassessment depends on observed early-stage data, there is a risk that the error rates across stages cannot be maintained at the intended levels and subsequent work will investigate avoiding this inflation.
The proposed methodology may also be extended to enable mid-course adaptations. For instance, the basket trial will begin with a few subsets of interest, and then restrict enrollment to the ones wherein patients benefit satisfactorily from the treatment based on an interim analysis. Boundaries for early stopping of certain subsets must be carefully defined to protect the overall error rates. With a reduction in the number of subsets, the synthesis weights $p_{qk}$ should be updated to satisfy the constraint of $\sum_q p_{qk} = 1$ for the late stage(s).

\section*{Software}

All statistical computing and analyses were performed using the software environment R version 4.0.3. Programming code for implementing the sample size formulae and reproducing the numerical results, is available at \url{https://github.com/haiyanzheng/BasketTrialsSSD}.

\section*{Acknowledgment}

This work was supported by Cancer Research UK through Dr Zheng's Population Research Postdoctoral Fellowship {\color{blue} (}RCCPDF\textbackslash 100008{\color{blue} )}.
Prof Jaki and Dr Mozgunov received funding from the UK Medical Research Council (MC\rule{2mm}{0.4pt}UU\rule{2mm}{0.4pt}00002/14). This report is independent research supported by the National Institute for Health Research (NIHR) Cambridge Biomedical Research Centre (BRC-1215-20014) and the NIHR Advanced Fellowship (Dr Mozgunov, NIHR300576). The views expressed in this publication are those of the authors and not necessarily those of the NHS, the NIHR or the Department of Health and Social Care (DHSC).\\

\noindent {\it Conflict of Interest}: None declared.


\bibliographystyle{apalike}



\clearpage

\renewcommand\thefigure{\arabic{figure}}    
\setcounter{figure}{0}

\begin{figure}[!ht]
\centering
\includegraphics[width=1.05\linewidth]{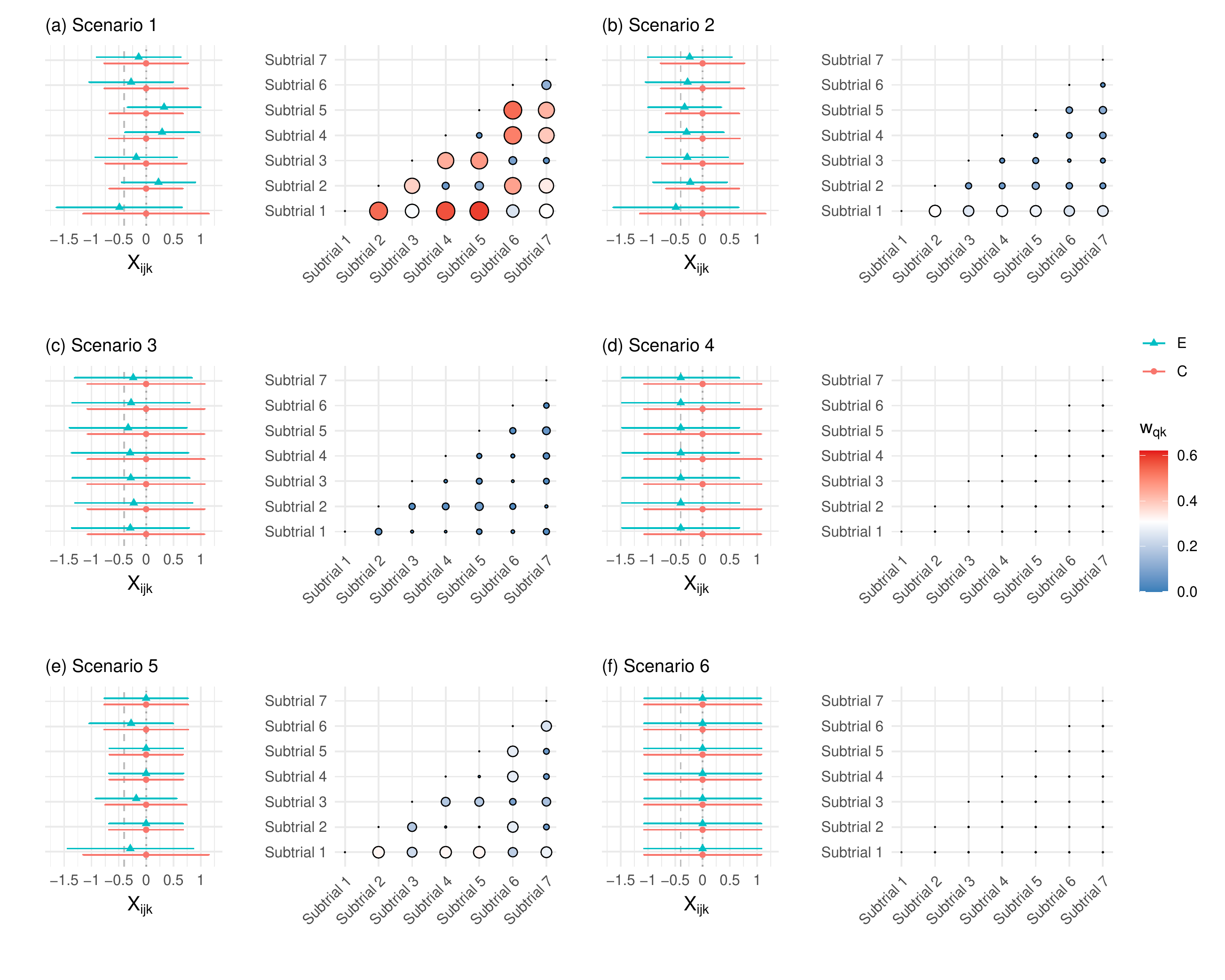}
\caption{Simulation scenarios depicted by the distributions of $X_{ijk}, j = E, C, \, k = 1, \dots, 7$, along with the visualisation of $w_{qk}$ for pairwise discounting of incommensurate complementary subtrial data. The green triangle and red circle mark the normal means $\mu_{Ek}$ and $\mu_{Ck}$, respectively; the endpoints of each horizontal bar are computed by $\mu_{jk}\pm 1.96\sigma_k^2$. The dashed and dotted vertical lines indicate the effect size of $\delta$ and naught, respectively.}
\label{fig:SimSc}
\end{figure}

\newpage

\begin{table}[!ht]
\centering
\caption{The required subtrial sample sizes, computed using the proposed methodology or the approach of no borrowing, to detect a difference of $\delta = -0.4$, when setting $\eta = 0.95, \zeta = 0.80$.}
\begin{tabular}{@{\extracolsep{2.0pt}}llcccccccc@{}}
\toprule
& &\multicolumn{7}{@{}c}{Required subtrial sample size}    \\
  \cline{3-9}   \\[-0.4em]
& & $k = 1$ & $k = 2$  & $k = 3$ & $k = 4$  & $k = 5$  &  $k = 6$  & $k = 7$  & Total \\
\midrule
\multirow{2}{*}{Scenario 1}  & $n_k^0$ (No borrowing) & 53.2  & 18.4  & 22.3  & 18.6  & 18.3  & 23.7  & 23.7  & 178.2 \\
& $n_k$ (Proposed) & 52.0  & 17.3  & 20.5  & 17.0  & 17.1  & 22.5  & 22.0 & 168.4 \\[0.2em]
   \cline{1-10}  \\[-0.4em] 
\multirow{2}{*}{Scenario 2} 
& $n_k^0$  (No borrowing) & 53.2  & 18.4  & 22.3  & 18.6  & 18.3  & 23.7  & 23.7  & 178.2  \\
& $n_k$ (Proposed) & 50.6  & 15.7  & 17.4  & 15.1  & 15.6  & 18.9  & 19.6  & 152.9 \\[0.2em]
   \cline{1-10}  \\[-0.4em] 
\multirow{2}{*}{Scenario 3} & $n_k^0$ (No borrowing)  & 46.4  & 46.4  & 46.4  & 46.4  & 46.4  & 46.4  & 46.4  & 324.8  \\ 
& $n_k$ (Proposed) & 23.3  & 32.0  & 22.6  & 24.4  & 32.9  & 23.3  & 30.2  & 188.7 \\[0.2em]
   \cline{1-10}  \\[-0.4em] 
\multirow{2}{*}{Scenario 4} & $n_k^0$  (No borrowing) & 46.4  & 46.4  & 46.4  & 46.4  & 46.4  & 46.4  & 46.4  & 324.8 \\
& $n_k$ (Proposed) & \, 8.9  & \, 8.9  & \, 8.9  & \, 8.9  & \, 8.9  & \, 8.9  & \, 8.9  & \, 62.3 \\[0.2em]
   \cline{1-10}  \\[-0.4em] 
\multirow{2}{*}{Scenario 5} & $n_k^0$ (No borrowing)  & 53.2  & 18.4  & 22.3  & 18.6  & 18.3  & 23.7  & 23.7  & 178.2 \\
& $n_k$ (Proposed) & 50.8  & 14.3  & 20.4  & 14.5  & 14.2  & 22.1  & 20.7  & 157.0 \\[0.2em]
   \cline{1-10}  \\[-0.4em] 
\multirow{2}{*}{Scenario 6}  & $n_k^0$ (No borrowing)  & 46.4  & 46.4  & 46.4  & 46.4  & 46.4  & 46.4  & 46.4   & 324.8   \\
& $n_k$ (Proposed) & \, 8.9  & \, 8.9  & \, 8.9  & \, 8.9  & \, 8.9  & \, 8.9  & \, 8.9  & \, 62.3 \\[0.2em]
\bottomrule
\end{tabular}
\label{tab:SimSS}
\end{table}

\newpage

\begin{figure}[!ht]
\centering
\includegraphics[width=1.05\linewidth]{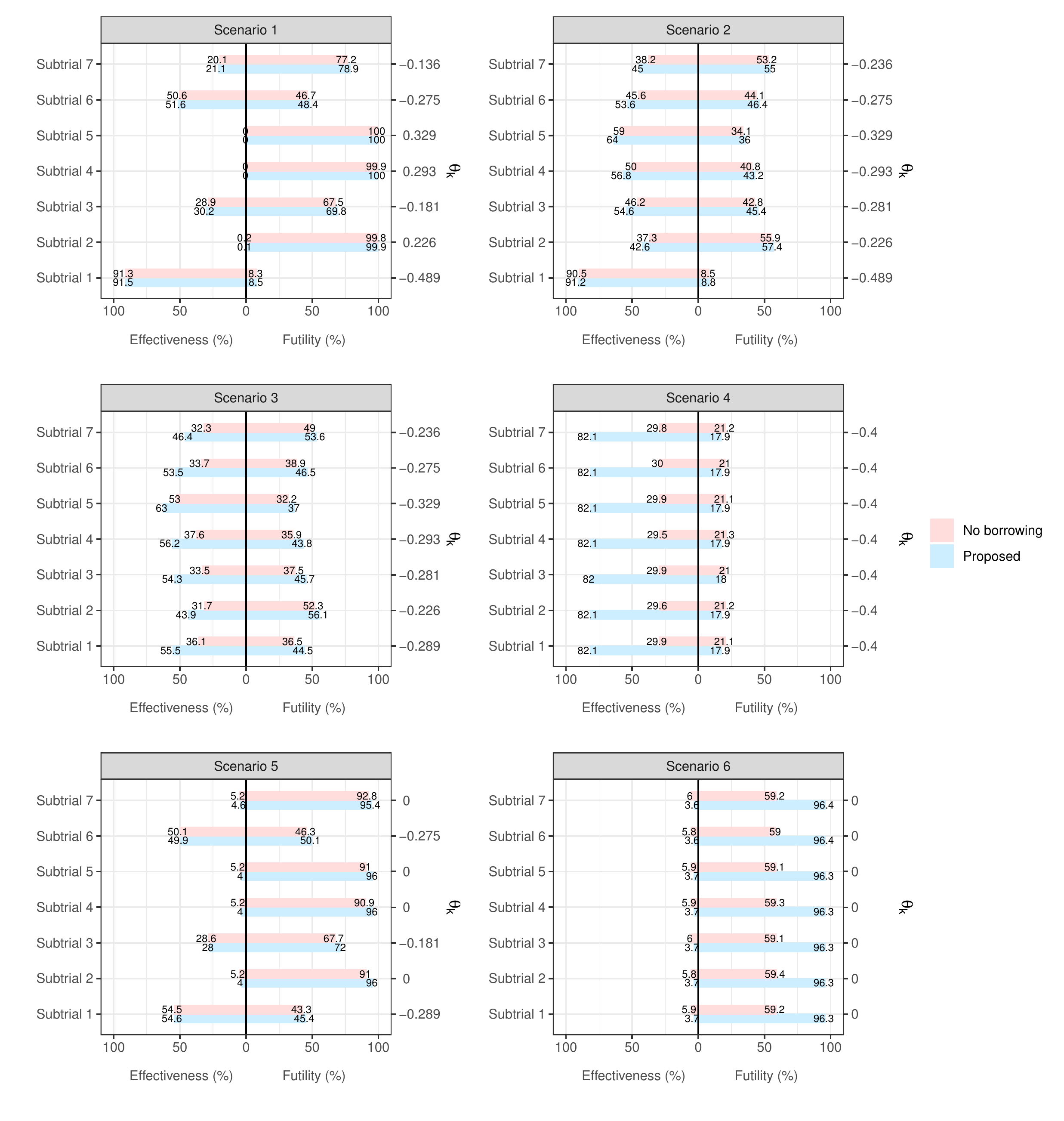}
\caption{Percentage of (sub)trials that conclude $E$ is efficacious (the left half of each plot) or not better than $C$ by $\delta = -0.4$, i.e., observing a shrinkage of 40\% in the tumour volume (the right half of each plot). True subtrial-specific treatment effects, $\theta_k = \mu_{Ek} - \mu_{Ck}$, have been indicated in a second $y$-axis.}
\label{fig:SimRes}
\end{figure}

\newpage

\begin{figure}[!ht]
\centering
\includegraphics[width=0.75\linewidth]{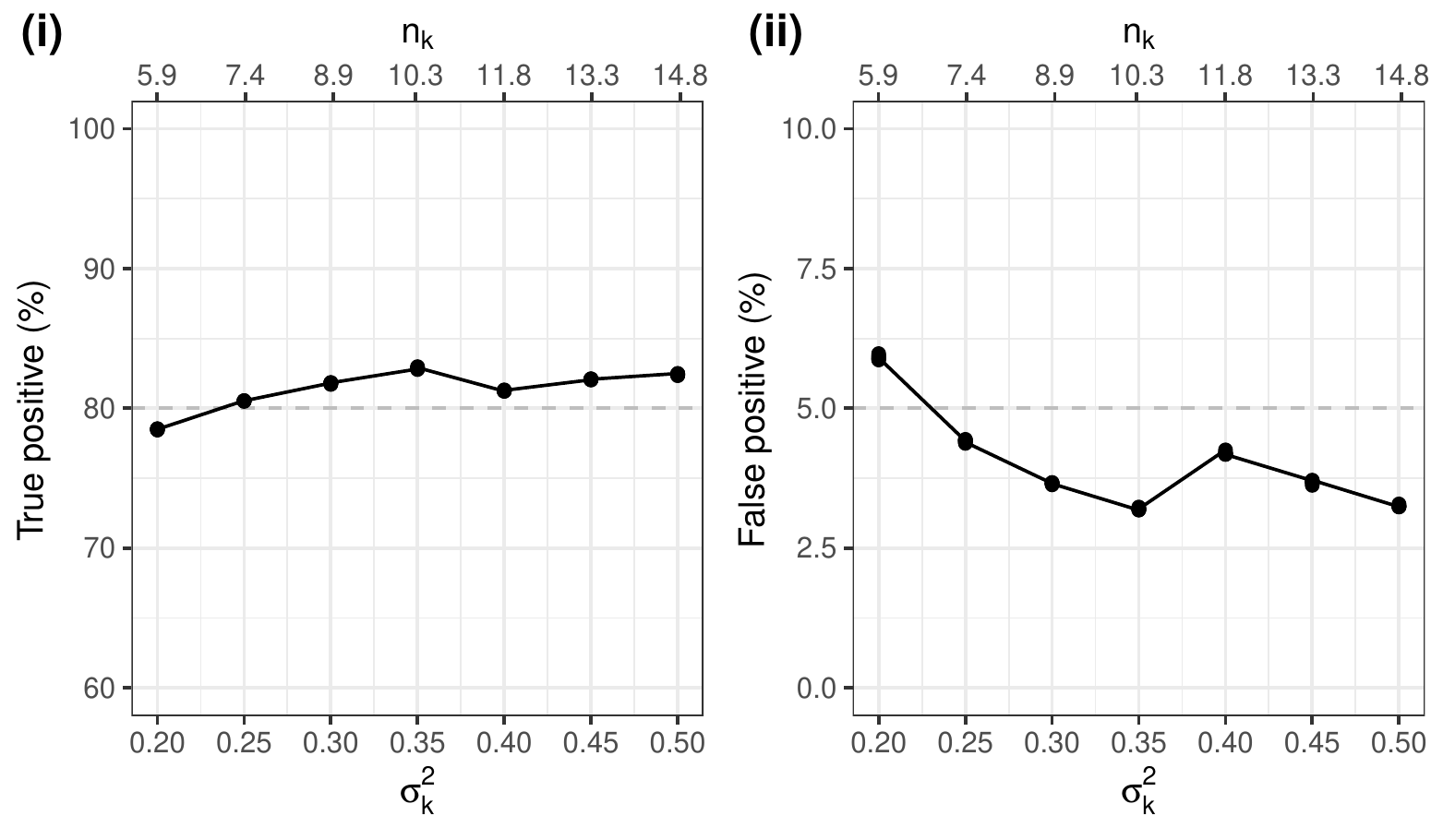}
\caption{The true and false positive rates per subtrials $k = 1, \dots, 7$, summarised based on the 100,000 replicates of basket trials, wherein $X_{iEk}$ are generated from $N(-0.4, \sigma_k^2)$ and $N(0, \sigma_k^2)$ distributions for panels (i) and (ii), respectively. The dashed lines indicate the pre-specified level of $\zeta = 0.80$ or $1-\eta = 0.05$.}
\label{fig:TPFP}
\end{figure}

\clearpage



\section*{Supplementary material}

# Supplementary Materials for:
# Bayesian sample size determination in basket trials borrowing information between subsets

**by Haiyan Zheng, Michael Grayling, Pavel Mozgunov, Thomas Jaki, James Wason**

**A. IMPACT OF SEVERAL KEY PARAMETERS ON THE SAMPLE SIZES**

In this section, we illustrate how sample sizes change along with certain key parameters, such as the variances and $w_{qk}$, for a special case, where the basket trial has two subgroups only (i.e., $K = 2$). For illustration, we have set $s_{0k}^2 = 100$, $a_1 = b_1 = 2$, $a_2 = 54, b_2 = 3$ and $c_0 = 0.05$. Assuming $\sigma_1^2 = 0.587^2$, $\sigma_2^2 = 0.345^2$ (values were extracted from the SUMMIT trial which appears in Section 3.2 of the main paper) and $\delta = 0.4$, sample sizes are sought for the inferences about $\theta_k$ with $\eta = 0.95$ and $\zeta = 0.80$. As expected, the variance $\sigma_k^2$ is a key factor for the determination of the corresponding $n_k$. More precisely, sample size of subtrial 1 is generally greater than that of subtrial 2 when holding other parameters fixed.

Figure S1 confirms that the sample size of subtrial 1 (subtrial 2) additionally relies on $w_{21}$ ($w_{12}$) which controls the degree of borrowing from subtrial 2 (subtrial 1): the $n_k$s in the same row within the subtrial 1 plot and those in the same column within the subtrial 2 plot remain equal by nearest integer for most cases. The only exceptions are the row with $w_{21} = 0$ in the subtrial 1 plot and the column with $w_{12} = 0$ in the subtrial 2 plot, where such variation is often trivial. We note this is caused by the use of Newton's method for solving the nonlinear equations. Figure S1 also indicates that an increase or decrease in $w_{qk}$ does not lead to linear change on the sample size of subtrial $k$, but with [0, 0.3] being a more sensitive range than (0.3, 1]. Substantial saving in the total sample size would be achieved if setting $w_{21}$ and $w_{12}$ to small values, for example, both as 0. Focusing on the third plane, the number at the bottom left goes larger when moving towards the top right.

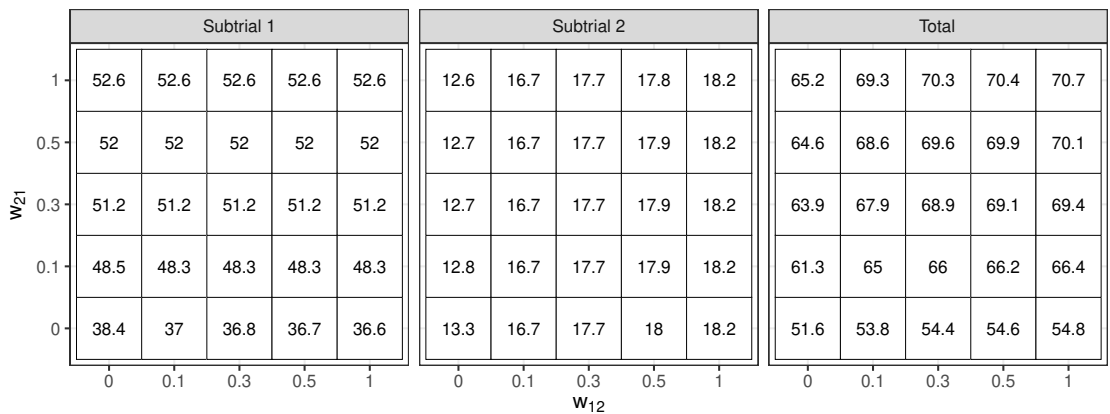

Figure S1: Bayesian sample sizes of the respective subtrials and entire basket trial with $K = 2$, given various levels of incommensurability.

Furthermore, $n_1$ and $n_2$ computed here based on the proposed sample size formula are bounded

by 52.6 (resulting from $w_{21} = 1$) and 18.2 (resulting from $w_{12} = 1$), respectively. These are close to $n_1^0 = 53.2$ and $n_2^0 = 18.4$, as obtained from the approach of no borrowing.

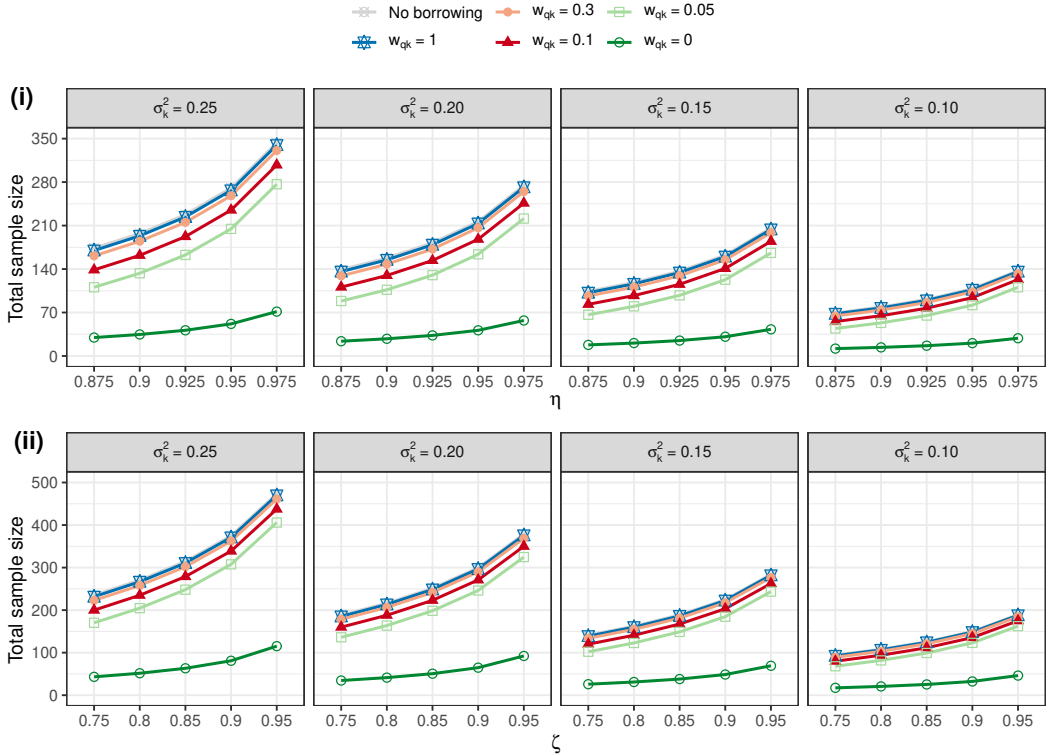

Figure S2: Total sample size targeting different levels of $\eta$ and $\zeta$. In panel (i), $\zeta$ is fixed at 0.80; while in panel (ii), $\eta$ is fixed at 0.95.

We now explore the behaviour of our Bayesian sample size formulae (when setting $0 \leq w_{qk} \leq 1$) in comparison to that of the approach of no borrowing. Focusing on homoskedastic cases (i.e., $\sigma_k^2$ remain identical across subtrials), Figure S2 shows that the trial sample size increases as the probability thresholds, $\eta$ or $\zeta$, increase. This is unsurprising, since a larger probability threshold would mean a more informative posterior distribution for $\theta_k$ to reach a decisive conclusion.

We also observe that sample size reduction is possible by setting a small value for $w_{qk}$. More specifically, $\sum n_k$ drops quickly when reducing $w_{qk}$ from 0.1 to 0 in both panels (i) and (ii). Looking across the plots of the same panel, the larger (smaller) the variances, the greater (less) the sample size would be required.

We then visualise the required sample sizes for different standardised effect sizes, $\delta/\sigma_k^2$, in Figure S3. We set $\sigma_k^2 = 0.25$ throughout this illustration and $\delta = 0.60, 0.55, 0.50, 0.45, 0.40, 0.35, 0.30$. Unsurprisingly, larger sample sizes are required to detect a smaller standardised effective size. Figure S3 also suggest that a value close to 0 for $w_{qk}$ results in substantial saving of sample size.

## B. SIMULATION SCENARIOS AND ADDITIONAL RESULTS

Table S1 lists the six scenarios that have been used in our simulation study of the main paper. Additional simulation results based on a sensitivity analysis will also be presented in this section.

Additional simulations have been performed to illustrate the proposed methodology controls

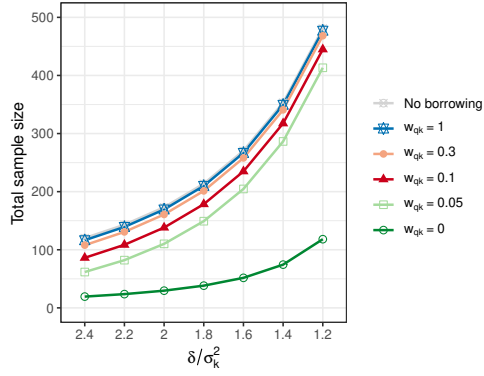

Figure S3: Total sample size, as a function of the standardised effect size, $\delta/\sigma_k^2$, by the proposed methodology and the approach of no borrowing, respectively, varying $w_{qk} = 0, 0.05, 0.1, 0.3, 1$.

Table S1: Simulation scenarios depicted as outcome distributions for the experimental treatment, $N(\mu_{Ek}, \sigma_k^2)$, $k = 1, \ldots, 7$. The corresponding outcome distribution for the control is $N(0, \sigma_k^2)$.

|  |  | Subtrial |  |  |  |  |  |  |
|---|---|---|---|---|---|---|---|---|
|  |  | $k=1$ | $k=2$ | $k=3$ | $k=4$ | $k=5$ | $k=6$ | $k=7$ |
| Scenario 1 | $\mu_{Ek}$ | -0.489 | 0.226 | -0.181 | 0.293 | 0.329 | -0.275 | -0.136 |
|  | $\sigma_k^2$ | 0.345 | 0.119 | 0.144 | 0.120 | 0.118 | 0.154 | 0.154 |
| Scenario 2 | $\mu_{Ek}$ | -0.289 | -0.226 | -0.281 | -0.293 | -0.329 | -0.275 | -0.236 |
|  | $\sigma_k^2$ | 0.345 | 0.119 | 0.144 | 0.120 | 0.118 | 0.154 | 0.154 |
| Scenario 3 | $\mu_{Ek}$ | -0.289 | -0.226 | -0.281 | -0.293 | -0.329 | -0.275 | -0.236 |
|  | $\sigma_k^2$ | 0.300 | 0.300 | 0.300 | 0.300 | 0.300 | 0.300 | 0.300 |
| Scenario 4 | $\mu_{Ek}$ | -0.400 | -0.400 | -0.400 | -0.400 | -0.400 | -0.400 | -0.400 |
|  | $\sigma_k^2$ | 0.300 | 0.300 | 0.300 | 0.300 | 0.300 | 0.300 | 0.300 |
| Scenario 5 | $\mu_{Ek}$ | -0.289 | 0 | -0.181 | 0 | 0 | -0.275 | 0 |
|  | $\sigma_k^2$ | 0.345 | 0.119 | 0.144 | 0.120 | 0.118 | 0.154 | 0.154 |
| Scenario 6 | $\mu_{Ek}$ | 0 | 0 | 0 | 0 | 0 | 0 | 0 |
|  | $\sigma_k^2$ | 0.300 | 0.300 | 0.300 | 0.300 | 0.300 | 0.300 | 0.300 |

the error rates at desired levels for each subtrial. Figure S4 visualises the accuracy of decision under two new mixed null scenarios, i.e., Scenario A with $\theta_1 = \theta_3 = \theta_6 = -0.4$ and Scenario B with $\theta_1 = \theta_3 = \theta_6 = -0.5$ along with $\theta_2 = \theta_4 = \theta_5 = \theta_7 = 0$ in both. The variances are set the same as those displayed in Scenario 5 above. As we can read from the figure, assuming the desired effect size $\delta = -0.4$, the percentage of simulations incorrectly declaring $E$ as efficacious for subtrials with $\theta_k = 0$ is maintained below 5% using the proposed methodology.

## C. EXTENDED APPLICATION TO USING A BINARY OUTCOME

When a binary outcome is used in randomised basket trials, the primary interest is to compare the response rates on the respective treatments, denoted by $\rho_{Ek}$ and $\rho_{Ck}$, in each subtrial $k = 1, \ldots, K$. In this section, we adapt the proposed methodology to enable borrowing of information in terms of log-odds ratios, which are approximately normally distributed [1].

Let $n_k$ be the subtrial sample size, for $k = 1, \ldots, K$. The log-odds ratio based on the individual

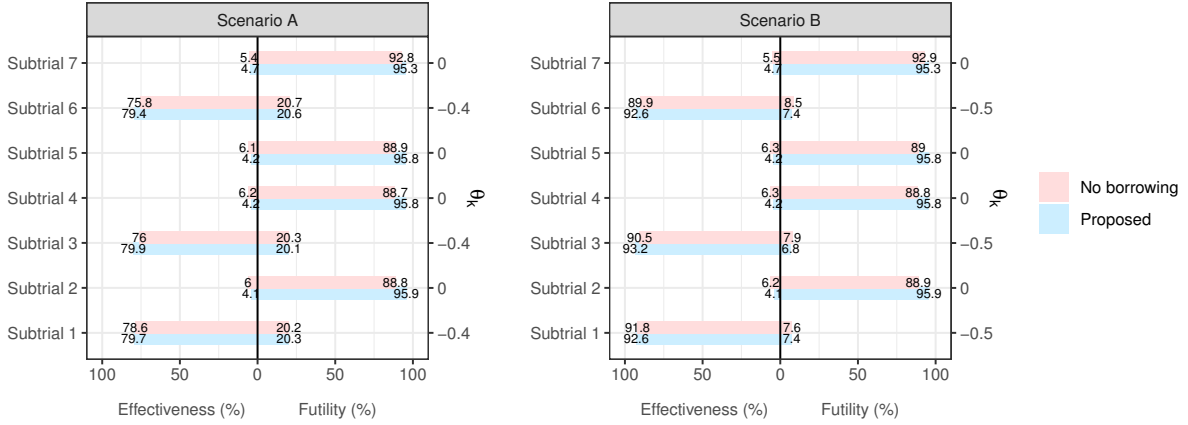

Figure S4: Percentage of (sub)trials that conclude $E$ is efficacious (the left half of each plot) or not better than $C$ by $\delta = -0.4$ under two new scenarios.

subtrial data is

$$\log(\hat{OR}_k) \dot\sim N\left(\log\left(\frac{\rho_{Ek}(1-\rho_{Ck})}{(1-\rho_{Ek})\rho_{Ck}}\right), \frac{1}{n_k}\left(\frac{1}{\rho_{Ek}} + \frac{1}{1-\rho_{Ek}} + \frac{1}{\rho_{Ck}} + \frac{1}{1-\rho_{Ck}}\right)\right).$$

We further let $\theta_k = \log[(\rho_{Ek}(1-\rho_{Ck}))/((1-\rho_{Ek})\rho_{Ck})]$. Following the methodology proposed in the main paper, we represent the complementary subtrial data in commensurate priors and place a two-component Gamma mixture prior on each commensurate parameter. This gives

$$\theta_k \mid x_k, x_{(-k)} \dot\sim N\left(d_{\theta_k}, \left(\frac{1}{\sum_q p_{qk}^2 \xi_{qk}^2} + n_k \left(\frac{1}{\rho_{Ek}(1-\rho_{Ek})} + \frac{1}{\rho_{Ck}(1-\rho_{Ck})}\right)^{-1}\right)^{-1}\right), \quad \text{(S1)}$$

with

$$\xi_{qk}^2 = \left(\frac{1}{s_{0q}^2} + n_q \left(\frac{1}{\rho_{Eq}(1-\rho_{Eq})} + \frac{1}{\rho_{Cq}(1-\rho_{Cq})}\right)^{-1}\right)^{-1} + \frac{w_{qk} b_1}{a_1 - 1} + \frac{(1 - w_{qk}) b_2}{a_2 - 1}.$$

Applying the same decision criterion with a clinically meaningful effect size, denoted by $\delta$, the subtrial sample sizes can be found so that the $K$ nonlinear equations hold simultaneously:

$$n_k \geq \left[\left(\frac{z_\eta + z_\zeta}{\delta}\right)^2 - \frac{1}{\sum_q p_{qk}^2 \xi_{qk}^2}\right] \left(\frac{1}{\rho_{Ek}(1-\rho_{Ek})} + \frac{1}{\rho_{Ck}(1-\rho_{Ck})}\right), \quad \forall k = 1, \ldots, K. \quad \text{(S2)}$$

### D. EXTENDED APPLICATION TO USING A TIME-TO-EVENT OUTCOME

We now consider extending the proposed sample size formulae to design randomised basket trials with a time-to-event outcome. For simplicity, we follow George and Desu [2] to assume that the event time, denoted by $T_{ijk}$, has an exponential distribution:

$$T_{ijk} \sim \text{Exp}(\pi_{jk}), \; i = 1, \ldots, n_k; \; j = E, C; \; k = 1, \ldots, K,$$

where the rate parameter $\pi_{jk} > 0$. Denote the average event time on treatment group $j$ by $\bar{T}_{jk}$ and the number of events by $D_{jk}$ in subtrial $k = 1, \ldots, K$. By the centrial limit theorem, we know

$$\bar{T}_{jk} \dot\sim N \left( \frac{1}{\pi_{jk}}, \frac{1}{D_{jk}\pi_{jk}} \right).$$

Using the delta method, we obtain

$$\log(\bar{T}_{jk}) \dot\sim N \left( -\log(\pi_{jk}), \frac{1}{D_{jk}} \right),$$

and further that

$$\log \left( \frac{\bar{T}_{Ek}}{\bar{T}_{Ck}} \right) \dot\sim N \left( \log(\frac{\pi_{Ck}}{\pi_{Ek}}), \frac{1}{D_{Ek}} + \frac{1}{D_{Ck}} \right).$$

Likewise, we let $\theta_k = \log(\frac{\pi_{Ck}}{\pi_{Ek}})$ and follow the proposed Bayesian methodology to enable borrowing of information between subtrials. This gives

$$\theta_k \mid x_k, x_{(-k)} \dot\sim N \left( d_{\theta_k}, \left( \frac{1}{\sum_q p_{qk}^2 \xi_{qk}^2} + \left( \frac{1}{D_{Ek}} + \frac{1}{D_{Ck}} \right)^{-1} \right)^{-1} \right), \quad \text{(S3)}$$

with

$$d_{\theta_k} = \frac{\frac{1}{D_{Ek}} + \frac{1}{D_{Ck}}}{\sum_q p_{qk}^2 \xi_{qk}^2 + \frac{1}{D_{Ek}} + \frac{1}{D_{Ck}}} \cdot \sum_q p_{qk} \lambda_{qk} + \frac{\sum_q p_{qk}^2 \xi_{qk}^2}{\sum_q p_{qk}^2 \xi_{qk}^2 + \frac{1}{D_{Ek}} + \frac{1}{D_{Ck}}} \cdot \log \left( \frac{\bar{t}_{Ek}}{\bar{t}_{Ck}} \right)$$

and

$$\xi_{qk}^2 = \left( \frac{1}{s_{0q}^2} + \left( \frac{1}{D_{Eq}} + \frac{1}{D_{Cq}} \right)^{-1} \right)^{-1} + \frac{w_{qk} b_1}{a_1 - 1} + \frac{(1 - w_{qk}) b_2}{a_2 - 1}.$$

Applying the same decision criterion as the main paper, the subtrial sample sizes can be found according to

$$\frac{D_{Ek} D_{Ck}}{D_{Ek} + D_{Ck}} \geq \left( \frac{z_\eta + z_\zeta}{\delta} \right)^2 - \frac{1}{\sum_q p_{qk}^2 \xi_{qk}^2}.$$

Equivalently,

$$D_k \geq \frac{1}{R_k(1 - R_k)} \left[ \left( \frac{z_\eta + z_\zeta}{\delta} \right)^2 - \frac{1}{\sum_q p_{qk}^2 \xi_{qk}^2} \right], \quad \text{(S4)}$$

where $D_k = D_{Ek} + D_{Ck}$ denotes the total number of events required per subtrial $k$, and $R_k$ is the randomisation ratio to the experimental treatment $E$.

### E. EXTENDED APPLICATION TO SINGLE-ARM SETTINGS WITH A BINARY OUTCOME

As one reviewer noted, basket trials are frequently conducted within a phase II oncology setting. Suppose that patients with a common feature (e.g., genomic aberration or clinical symptom) are enrolled to a basket trial with $K$ subsets; each subset represents a subtype of the disease. Within each trial subset (i.e., 'subtrial' for short) $k = 1, \ldots, K$, patients are treated by a new treatment (labelled $E$) and observed to be either a 'responder' or 'non-responder' according to predefined criteria relating to the tumour shrinkage. Let $Y_k$ denote the number of responders and $n_k$ the number of patients per subtrial $k$. Thus, $\mathbb{E}(Y) = n_k \rho_k$ and $\text{Var}(Y) = n_k \rho_k (1 - \rho_k)$, where $\rho_k$ is the

5

subtrial-specific response rate. By the delta method, the estimator of log-odds has an asymptotic normal distribution of

$$\log\left(\frac{\hat{\rho}_k}{1-\hat{\rho}_k}\right) \dot{\sim} N\left(\log\left(\frac{\rho_k}{1-\rho_k}\right), \frac{1}{n_k \rho_k(1-\rho_k)}\right)$$

Letting $\theta_k = \log(\rho_k/(1-\rho_k))$, we follow the proposed methodology to specify commensurate priors for $\theta_k$, based on the complementary subtrial data, and place a two-component Gamma mixture prior on each commensurate parameter. This then leads to

$$\theta_k \mid x_k, x_{(-k)} \dot{\sim} N\left(d_{\theta_k}, \left(\frac{1}{\sum_q p_{qk}^2 \xi_{qk}^2} + n_k \rho_k(1-\rho_k)\right)^{-1}\right), \tag{S5}$$

with

$$\xi_{qk}^2 = \left(\frac{1}{s_{0q}^2} + n_q \rho_q(1-\rho_q)\right)^{-1} + \frac{w_{qk} b_1}{a_1 - 1} + \frac{(1-w_{qk}) b_2}{a_2 - 1}.$$

Applying the same decision criterion, the subtrial sample sizes can be found to satisfy the $K$ non-linear equations simultaneously:

$$n_k \geq \frac{1}{\rho_k(1-\rho_k)} \left[\left(\frac{z_\eta + z_\zeta}{\delta}\right)^2 - \frac{1}{\sum_q p_{qk}^2 \xi_{qk}^2}\right]. \tag{S6}$$

## F. PRACTICAL IMPLEMENTATION OF THE PROPOSED SAMPLE SIZE FORMULAE

Figure S5 guides through the specification of parameters (specific to the design and data type) to implement the proposed Bayesian sample size formulae for basket trials, wherein borrowing of information is enabled wih $w_{qk} < 1$.

As noted, setting $\eta = 0.95$ and $\zeta = 0.80$ creates a resemblance to the frequentist formulation of the problem in situations of no borrowing (with all $w_{qk} = 1$). The user may modify these probability thresholds for their own case. Raising the level of $\eta$ or $\zeta$ would mean an increase in the sample size. We recommend the user to visualise such changes following the pattern of Figure S2. This would lead to effective communication with the stakeholder for finding sample sizes that are both affordable and sensible for estimating the effect size. In the randomised, two-arm settings, we have considered equal randomisation throughout by setting $R_1 = \cdots = R_K = 0.5$. The user may change the allocation ratio for their context.

Having chosen a design and an outcome to evaluate the clinical benefit, the user should seek information regarding (i) plausible response rate ($\rho_k$ or $\rho_{jk}$ for $j = E, C$, if using a binary outcome) or (ii) variability of responses ($\sigma_k^2$, if using a continuous outcome), both specific to subtrial $k = 1, \ldots, K$, along with a clinically meaningful effect size (denoted by $\delta'$, $\delta^*$, $\delta$ or $\delta^\dagger$ in Figure S5). This would be substantially easier if pilot data or historical information is available.

Specification of $w_{qk}$ to correctly reflect the pairwise incommensurability might be challenging in practice. This should best be informed by pilot data or existing information from relevant investigation. Following the derivation of the sample size formulae in Section 2 of the main paper, values for $w_{qk}$ can be chosen independently of the variances, $\sigma_k^2$, $k = 1, \ldots, K$. For continuous data, however, values of $\sigma_k^2$ and $w_{qk}$ can both be based on existing information from, e.g., a pilot study or preceding trial. We have exemplified this in Section 3.2 of the main paper, where the Hellinger distance is used to measure the discrepancy between the outcome distributions. The latter can

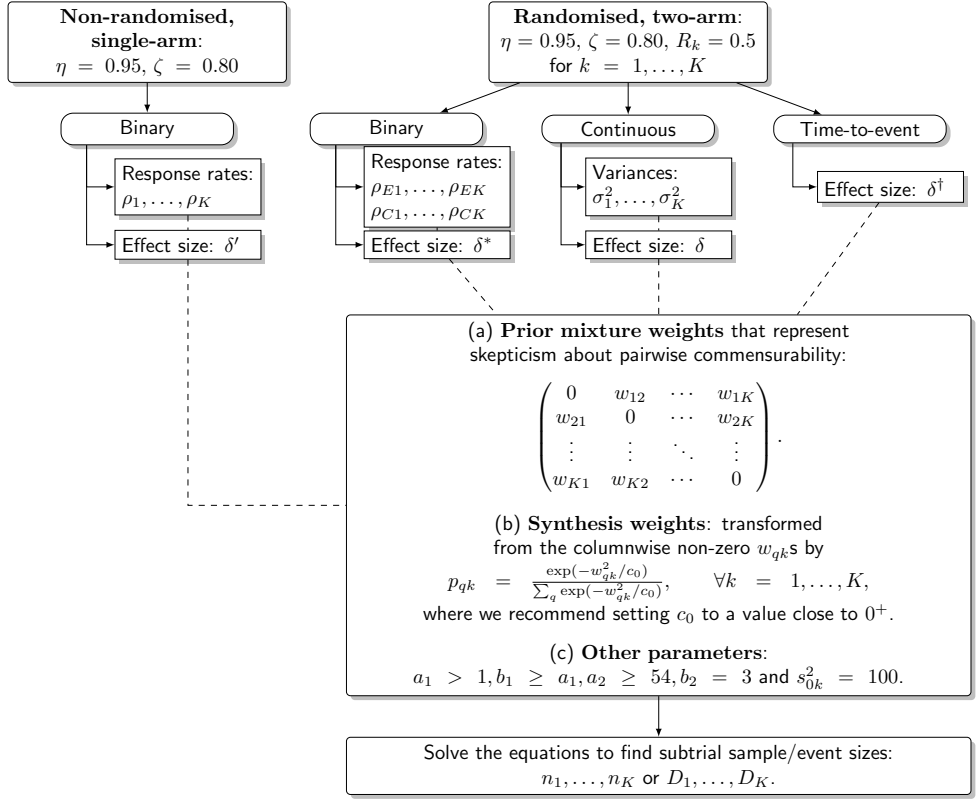

Figure S5: A roadmap for parameter specification in the proposed sample size formulae to design basket trials that may adopt a (i) non-randomised, single-arm, or (ii) randomised, two-arm setting in the respective subsets. Specific to the data type, the effect size (denoted by $\delta'$, $\delta^*$, $\delta$ or $\delta^\dagger$) is on a scale of log-odds, log-odds ratio, mean difference or log-hazard ratio, respectively.

be assumed with available data from a preceding trial (i.e., the SUMMIT trial in Section 3.2). For basket trials using a binary or time-to-event outcome, one may likewise compute the Hellinger distance between asymptotic normal distributions of the subtrial-specific log-odds, log-odds ratio or log-hazard ratio.

In these circumstances, the assumed values for response rates or variances affect the levels of $w_{qk}$, and the impact on the corresponding subtrial sample sizes may thus be of interest. We refer the reader to Table 1 of Section 4, which expands the illustration of sample size determination with various scenarios. Specifically, the variances in subsets 2 – 7 are increased from Scenario 1 (used in Section 3.2) to Scenario 3, which gives a different set of values for $w_{qk}$; see Figure 2 of the main paper for the visualisation. The resulting sample size is then an outcome of the increased variances and declined levels of pairwise incommensurability (i.e., increased amount of borrowing): the sample sizes $n_2, \ldots, n_7$ based on the proposed approach increase from Scenario 1 to Scenario 3, whereas the magnitude of such increase is not as large as that of $n_2^0, \ldots, n_7^0$ based on the approach of no borrowing; please see Table 1 of the main paper for this.

For the hyperparameters, the only constraint is $a_1, a_2 > 1$. As shown with illustrative examples in the main paper, we recommend setting these hyperparameters to give a small prior mean (e.g.,

## G. CONSEQUENCE OF PARAMETER MISSPECIFICATION FOR $w_{qk}$

Following the specification of $w_{qk}$ outlined in Section 3.2 of the main paper, we compute the Hellinger distance between any pair of $N(\mu_{Ek}, \sigma_k^2)$. Thus, we obtain

$$\begin{pmatrix} w_{11} & w_{12} & \cdots & w_{17} \\ w_{21} & w_{22} & \cdots & w_{27} \\ \vdots & \vdots & \cdots & \vdots \\ w_{71} & w_{72} & \cdots & w_{77} \end{pmatrix} = \begin{pmatrix} 0 & 0.539 & 0.300 & 0.571 & 0.591 & 0.246 & 0.312 \\ 0.539 & 0 & 0.384 & 0.068 & 0.105 & 0.457 & 0.342 \\ 0.300 & 0.384 & 0 & 0.439 & 0.470 & 0.087 & 0.044 \\ 0.571 & 0.068 & 0.439 & 0 & 0.037 & 0.508 & 0.397 \\ 0.591 & 0.105 & 0.470 & 0.037 & 0 & 0.537 & 0.429 \\ 0.246 & 0.457 & 0.087 & 0.508 & 0.537 & 0 & 0.125 \\ 0.312 & 0.342 & 0.044 & 0.397 & 0.429 & 0.125 & 0 \end{pmatrix},$$

for Scenario 1 and all $w_{qk} = 0$ for Scenario 6. When the same matrix of $w_{qk}$ is used to analyse the trial, the operating characteristics of the proposed design are as visualised in plots (a) and (f) of Figure 3 of the main paper, respectively. For our interest in the impact of misspecification of $w_{qk}$, we assume that three $7 \times 7$ matrix, wherein $w_{qk} = 0.1, 0.3, 0.5$ (for $q \neq k$) along with $w_{kk} = 0$, $k = 1, \ldots, 7$, would be applied in the Bayesian analysis. In the following sensitivity analysis, we simulate 100,000 replicates of the basket trial with the same parameter configuration as that of the main simulation study.

Table S2: Percentage of (sub)trials that conclude effectiveness or futility of $E$ in situations where the values of $w_{qk}$ are correctly or incorrectly specified.

|  |  |  | Subtrial | | | | | | |
| --- | --- | --- | --- | --- | --- | --- | --- | --- | --- |
|  |  |  | $k=1$ | $k=2$ | $k=3$ | $k=4$ | $k=5$ | $k=6$ | $k=7$ |
| Scenario 1 | True $w_{qk}$ | Effectiveness (%) | 91.5 | 0.1 | 30.2 | 0 | 0 | 51.6 | 21.1 |
|  |  | Futility (%) | 8.5 | 99.9 | 69.8 | 100 | 100 | 48.4 | 78.9 |
|  | $w_{qk} = 0.1$ | Effectiveness (%) | 88.8 | 0.2 | 25.6 | 0 | 0 | 46.1 | 17.5 |
|  |  | Futility (%) | 14.0 | 99.9 | 76.4 | 100 | 100 | 57.8 | 84.3 |
|  | $w_{qk} = 0.3$ | Effectiveness (%) | 90.5 | 0.2 | 27.7 | 0 | 0 | 49.0 | 19.1 |
|  |  | Futility (%) | 10.0 | 99.8 | 70.9 | 99.9 | 100 | 50.6 | 79.9 |
|  | $w_{qk} = 0.5$ | Effectiveness (%) | 90.8 | 0.2 | 28.2 | 0 | 0 | 49.7 | 19.5 |
|  |  | Futility (%) | 9.3 | 99.8 | 69.5 | 99.9 | 100 | 49.1 | 78.9 |
| Scenario 6 | True $w_{qk}$ | Effectiveness (%) | 3.7 | 3.7 | 3.7 | 3.7 | 3.7 | 3.6 | 3.6 |
|  |  | Futility (%) | 96.3 | 96.3 | 96.3 | 96.3 | 96.3 | 96.4 | 96.4 |
|  | $w_{qk} = 0.1$ | Effectiveness (%) | 2.6 | 2.6 | 2.7 | 2.7 | 2.7 | 2.5 | 2.7 |
|  |  | Futility (%) | 73.8 | 74.2 | 73.8 | 74.1 | 73.8 | 73.8 | 73.8 |
|  | $w_{qk} = 0.3$ | Effectiveness (%) | 4.2 | 4.1 | 4.3 | 4.2 | 4.2 | 4.1 | 4.3 |
|  |  | Futility (%) | 64.8 | 65.1 | 64.9 | 65.1 | 64.8 | 64.7 | 64.8 |
|  | $w_{qk} = 0.5$ | Effectiveness (%) | 4.7 | 4.7 | 4.8 | 4.8 | 4.7 | 4.7 | 4.8 |
|  |  | Futility (%) | 62.6 | 62.9 | 62.7 | 62.9 | 62.7 | 62.5 | 62.6 |

Table S2 shows that when true values of $w_{qk}$ are used in the analysis, all of the simulated trials will be assigned a decisive subtrialwise conclusion (100% = Effectiveness% + Futility%). When $w_{qk}$ is specified as a greater value in the analysis than that used in the design (meaning that the amount of borrowing is attenuated), a smaller proportion of trials will reach a decisive decision


making. This is evident by the results for Scenario 6 in Table S2 at the rows of $w_{qk} = 0.1, 0.3, 0.5$ (all being $> 0$ which is the value of $w_{qk}$ used in the design). The more the values deviate from the designated level, the fewer trials could be concluded correctly on the futility of treatment in Scenario 6. By contrast, when $w_{qk}$ would be specified to a smaller value, some trials may produce an ambiguous decision on effectiveness or futility; see, for example, $k = 1$ in Scenario 1 with all $w_{qk} = 0.1$ which are consistently lower than the values used for the design. When the magnitude of deviation is small, the impact of misspecification is trivial: the percentages for $k = 7$ in Scenario 1 with $w_{qk} = 0.3$ are comparable to those with true $w_{qk}$ used in the analysis.

Finally, we would reemphasise that the change of $w_{qk}$ does not lead to a linear increase or decrease in the corresponding sample size. As Figure 1 in the main paper illustrates, the region bounded by 0 and 0.3 tends to be more sensitive to the other half, particularly from 0.5 to 1. Therefore, attention would be needed for the misspecification of $w_{qk}$, if the true values are *{either}* *within the sensitive region {or} the deviated across the region* (e.g., setting $w_{qk} = 0.55$ in the design yet $w_{qk} = 0.25$ in the analysis, or the other way around).




\end{document}